\documentstyle[preprint,aps,psfig]{revtex}
\newcommand{\beq}{\begin{equation}}
\newcommand{\eeq}{\end{equation}} 
\newcommand{\beqa}{\begin{eqnarray}}
\newcommand{\eeqa}{\end{eqnarray}} 
\newcommand{\ba}{\begin{array}} 
\newcommand{\ea}{\end{array}} 

\begin{document}
\draft

\widetext 
\title{Parametric Resonant Phenomena in Bose-Einstein Condensates: \\ 
Breaking of Macroscopic Quantum Self-Trapping} 
\author{Luca Salasnich} 
\address{Istituto Nazionale per la Fisica della Materia, 
Unit\`a di Milano, \\
Dipartimento di Fisica, Universit\`a di Milano, \\
Via Celoria 16, 20133 Milano, Italy \\
E-mail: salasnich@mi.infm.it} 
\maketitle 
\begin{abstract} 
We analyze the periodic tunneling of a Bose-Einstein 
condensate in a double-well potential which has an oscillating 
energy barrier. We show that the dynamics of the Bose condensate critically 
depends on the frequency $\omega$ of the oscillating energy barrier. 
In the regime of periodic macroscopic quantum tunneling (PMQT) 
with frequency $\omega_J$, the population imbalance of the condensate in the 
two wells can be enhanced 
under the condition of parametric resonance $\omega = 2 \omega_J$. 
Instead, in the regime of macroscopic quantum self-trapping (MQST),  
we find that MQST can be reduced or suppressed under the condition 
of parametric resonance between 
the frequency $\omega$ of the energy barrier and the frequency $\omega_{ST}$ 
of oscillation through the barrier of the very small 
fraction of particles which remain untrapped during MQST. 
\end{abstract}

\vskip 1. truecm

\narrowtext

\newpage 

\section{Introduction}

Macroscopic quantum tunneling 
with dilute Bose-Einstein condensates of alkali-metal atoms 
has been the subject of many theoretical [1-4] and experimental 
[5-7] studies. 
Here we consider the problem of trapped Bosons 
in a double-well potential and investigate the effect of 
a periodically varying barrier in the tunneling of the Bose condensate. 
We study the problem using the two-mode classical-like 
equations [1,2] and find that remarkable effects are related to 
parametric resonance [8] both in the periodic macroscopic quantum 
tunneling regime and in the macroscopic quantum self-trapping regime. 

\section{Tunneling in a double-well potential} 

In a dilute gas of $N$ Bosons at zero temperature practically 
all particles are in the same single-particle state of the 
density matrix $\rho({\bf r},{\bf r}';t)$ [9]. This macroscopically occupied 
single-particle state is called Bose-Einstein condensate and 
its wavefunction $\psi({\bf r},t)$ is well described by the 
Gross-Pitaevskii equation 
\beq 
i\hbar {\partial \over \partial t}\psi({\bf r},t) = 
\left[ -{\hbar^2\over 2m} \nabla^2 
+ U({\bf r}) + g |\psi({\bf r},t)|^2 \right] \psi({\bf r},t)  \; , 
\eeq 
where $U({\bf r})$ is the external potential, $g=4\pi\hbar^2a_s/m$ 
is the inter-atomic strength with $a_s$ the s-wave 
scattering length, and the wavefunction is normalized to $N$ [10]. 
\par  
Let us consider a double-well external potential $U({\bf r})$. 
In the tunneling regime, the wavefunction of the Bose-condensate 
can be approximated in the following way 
\beq 
\psi({\bf r},t) = \sqrt{N_1(t)}e^{i\phi_1(t)} \psi_1({\bf r}) + 
\sqrt{N_2(t)} e^{i\phi_2(t)} \psi_2({\bf r}) \; , 
\eeq 
where $N_1(t)$ is the number of particles in the first well and 
$N_2(t)$ is the number of particles in the second well, such as 
$N=N_1(t)+N_2(t)$ [1]. If the double-well is simmetric, 
it is not difficult to show 
that the time-dependent behavior of the condensate in the tunneling 
energy range can be described, with a suitable rescaling of time, 
by the two-mode equations 
\beq
{\dot \zeta}=-\sqrt{1-\zeta^2}\sin{\phi} \; , \;\;\;\;\;\;\; 
{\dot \phi}=\Lambda \zeta +{\zeta\over \sqrt{1-\zeta^2}}\cos{\phi} \; ,
\eeq 
where $\zeta=(N_1-N_2)/N$ is the fractional population 
imbalance of the condensate in the two wells, 
$\phi=\phi_1-\phi_2$ is the relative phase 
(which can be initially zero), and $\Lambda = 2 E_I/E_T$ with 
$E_I$ the inter-atomic energy and 
$E_T$ the tunneling energy, i.e. the kinetic+potential energy 
splitting between the ground state and the quasi-degenerate 
odd first excited state of the GP equation [1,2]. 
Note that $E_T=\hbar \omega_0$, 
where $\omega_0$ is the oscillation frequency of the Bose condensate 
between the two wells when the inter-atomic interaction 
is zero ($E_I=0$) [1-4]. 
\par 
The two-mode equations have been studied by many authors and 
they are the Hamilton equations of the following Hamiltonian 
\beq
H = {\Lambda\over 2}\zeta^2 - \sqrt{1-\zeta^2}\cos{\phi} \; , 
\eeq
where $\zeta$ is the conjugate momentum of the generalized 
coordinate $\phi$. Actually, it is possible to introduce [1] 
an equivalent Hamiltonian, given by 
\beq 
H'= {1\over 2} p_{\zeta}^2 + a \zeta^2 + b \zeta^4 \; ,  
\eeq
where 
\beq
a={1\over 2} \sqrt{1-\Lambda H_0} \; ,  \;\;\;\;\;\;\; 
b={\Lambda^2\over 8} \; ,  
\eeq 
with $H_0=\Lambda\zeta(0)^2/2-\sqrt{1-\zeta(0)^2}\cos{\phi(0)}$. 
For the Hamiltonian $H'$ the generalized coordinate is $\zeta$ and 
its conjugate momentum is $p_{\zeta}$. The equations of motion 
derived by $H'$ are 
\beq
{\dot \zeta} = p_{\zeta} \; ,  \;\;\;\;\;\;\; 
{\dot p_{\zeta}} = - 2 a \zeta - 4 b \zeta^3 \; . 
\eeq
\par 
The analysis of the two-mode equations (3) or (7) has shown that for 
$\Lambda <2$ there is a periodic macroscopic quantum tunneling (PMQT) 
of the Bose-condensate with Josephson-like oscillations 
of frequency $\omega_J = \omega_0 \sqrt{1+\Lambda}$ [1,2]. 
Instead, for $\Lambda >2$, there exists a critical 
$\zeta_c =2\sqrt{\Lambda -1}/\Lambda$ 
such that for $0 < \zeta(0) << \zeta_c$ there is PMQT  
of condensate with frequency $\omega_J$, 
but for $\zeta_c<\zeta(0) \leq 1$ there is macroscopic quantum 
self-trapping (MQST) of the condensate: even if the populations
of the two wells are initially set in an asymmetric state 
($\zeta(0)\ne 0$) they maintain, on the average, 
the original population imbalance with a very small periodic 
transfer of particles through the barrier with frequency  
$\omega_{ST}=\omega_0 \sqrt{2(\Lambda H_0-1)}$ [1,2]. 

\section{An oscillating barrier in the double-well potential} 

In this section we analyze the effect of a periodic 
oscillating energy barrier in the double-well potential. 
The presence of a oscillating barrier can be modelled 
by a time-dependent $\Lambda$. We choose the following form  
\beq 
\Lambda(t) = \Lambda_0 (1+\epsilon \sin{(\omega t)}) \; ,  
\eeq
where $\Lambda_0$ is the static value, $\epsilon$ is a small 
perturbation ($\epsilon \ll 1$) and $\omega$ is the oscillation 
frequency of the barrier. 
\par 
We numerically solve the two-mode equations (7) with (8) by using a 
forth-order Runge-Kutta algorithm. 
First we consider the case of pulsed macroscopic 
quantum tunneling (PMQT). We set $\Lambda_0 = 0.8$, $\zeta(0)=0.5$ 
and $\phi(0)=0$.  
In Figure 1 we plot the population imbalance 
$(t,\zeta)$ and its phase-space portrait $(\zeta , p_{\zeta})$ 
for some values of $\epsilon$ and $\omega$. 
For $\epsilon=0$ the population imbalance harmonically 
oscillates between the values $-0.5$ and $0.5$ 
with frequency $\omega_J=1.27$, but for $\epsilon\neq 0$ 
the motion is not fully harmonic. In particular, 
Figure 1 shows that under the parametric 
resonance condition $\omega = 2 \omega_J$ the dynamics of the 
population imbalance is modified: the amplitude of 
the oscillation is modulated and reaches the 
values $-1$ and $1$. 
\par 
Now we consider the case of macroscopic quantum 
self-trapping (MQST). We set $\Lambda_0 = 25$, $\zeta(0)=0.6$ 
and $\phi(0)=0$. Figure 2 confirms that for $\epsilon=0$ 
the population imbalance does not change sign and it 
periodically oscillates between the values $0.47$ 
and $0.6$ with frequency $\omega_{ST}=13.53$. 
As shown in Figure 2, MQST is strongly affected by 
the parametric resonance between 
the frequency $\omega_{ST}$ of MQST oscillation and the frequency 
$\omega$ of oscillation of the energy barrier. Moreover, 
Figure 3 shows that at the parametric resonance condition 
$\omega=2\omega_{ST}$ and with a sufficiently large perturbation 
($\epsilon =0.2$) the system eventually escapes from the self-trapping 
configuration. Note that this remarkable effect has been 
confirmed [11] by using the nonpolynomial Schr\"odinger equation, 
an effective one-dimensional equation derived from the three-dimensional 
Gross-Pitaevskii equation [12]. 

\section*{Conclusions} 

We have studied the periodic tunneling and the quantum self-trapping 
of a Bose-Einstein condensate in a double-well 
potential with an oscillating energy barrier. 
We have used the two-mode classical-like equations to find 
that the periodic macroscopic quantum tunneling can be enhanced and 
the macroscopic quantum self-trapping can be reduced or suppressed 
by changing the frequency of the oscillating barrier. 
In the latter case, we have proved that 
the system escapes from the self-trapping configuration 
if the the frequency of oscillation of the double-well energy barrier 
and the frequency of MQST oscillations of the condensate 
satisfy the parametric resonance condition 
with a sufficiently large perturbation of the energy barrier. 

\section*{References}

\begin{description}

\item{\ [1]} Smerzi, A., Fantoni, S., Giovannazzi, G. 
and Shenoy, S.R., 1997, Phys. Rev. Lett. {\bf 79} 4950;  
Raghavan, S., Smerzi, A., Fantoni, S., 
and Shenoy, S.R., 1999, Phys. Rev. A {\bf 59} 620. 

\item{\ [2]} Milburn, G.J., Corney, J., Wright, E., 
and Walls, D.F., 1997, Phys. Rev. A {\bf 55} 4318.  

\item{\ [3]} Zapata, I., Sols, F., and Leggett, A.J., 1998,  
Phys. Rev. A {\bf 57} R28. 

\item{\ [4]} Salasnich, L., Parola, A., and Reatto, L., 1999, 
Phys. Rev. A {\bf 60} 4171; 
Pozzi, B., Salasnich, L., Parola, A., and Reatto, L., 2000,     
Eur. Phys. J. D {\bf 11} 367. 

\item{\ [5]} Anderson, B.P. and Kasevich, M., 1999, 
Science {\bf 282} 1686. 

\item{\ [6]} Cataliotti, F.S., Burger, S., Fort, C., Maddaloni, P., 
Trombettoni, A., Smerzi, A., and M. Inguscio, 2001, 
Science {\bf 293} 843. 

\item{\ [7]} Morsch, O., Muller, J.H., Cristiani, M., Ciampini, D.,  
and Arimondo, E., 2001, Phys. Rev. Lett. {\bf 87} 140402. 

\item{\ [8]} Landau, L.D. and Lifsits, E.M., 1991,  
{\it Mechanics, Course of Theoretical Physics}, 
vol. 3 (Pergamon Press, Oxford); 
Arnold, V.I., 1990, {\it Mathematical Methods of Classical Mechanics} 
(Springer, Berlin). 

\item{\ [9]} Courteille, P.W., Bagnato, V.S., and 
Yukalov, Y.I., 2001, Laser Phys. {\bf 11}, 659. 

\item{\ [10]} Gross, E.P., 1961, Nuovo Cimento {\bf 20} 454; 
Pitaevskii, L.P., 1961, Zh. Eksp. Teor. Fiz. {\bf 40}, 646.

\item{\ [11]} Salasnich, L., Parola, A., and Reatto, L., 2002, 
J. Phys. B {\bf 35}, 3205. 

\item{\ [12]} Salasnich, L., 2002, Laser Phys. {\bf 12} 198; 
Salasnich, L., Parola, A., and Reatto, L., 2002, Phys. Rev. A 
{\bf 65} 043614. 

\end{description}

\newpage 

\begin{figure}
\centerline{\psfig{file=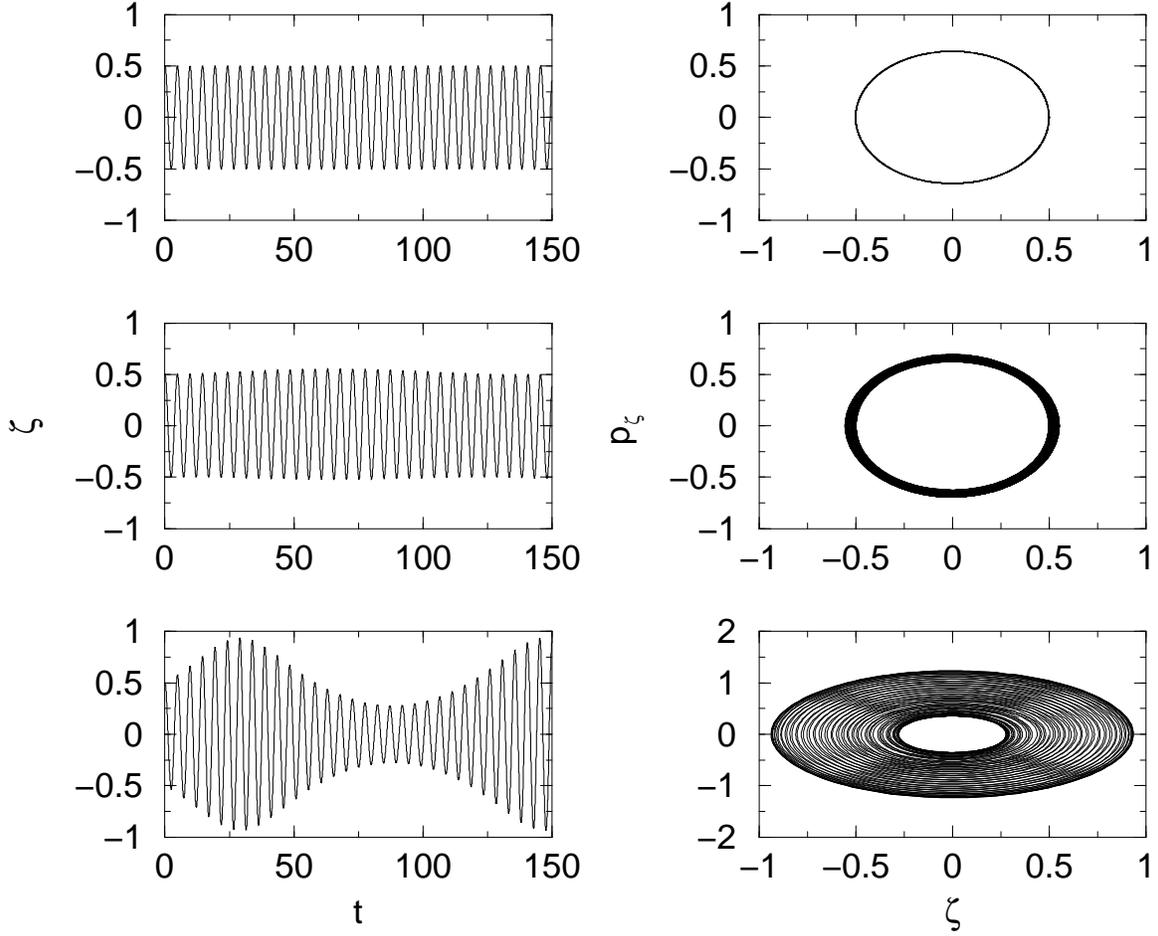,height=5.in}}
\caption{PMQT regime. Fractional population 
imbalance $\zeta(t)$ and its phase-space portrait 
$(\zeta,p_{\zeta})$. Initial conditions: 
$\zeta(0)=0.5$, $\phi(0)=0$. 
$\Lambda=\Lambda_0(1+\epsilon \sin{(\omega t)})$ 
where $\Lambda_0=0.8$. From top to bottom: 
(a) $\epsilon = 0$; (b) $\epsilon = 0.25$ and 
$\omega= \omega_{J}$; (c) $\epsilon =0.25$ and $\omega= 2 \omega_J$;  
where $\omega_J$ is the frequency of unperturbed PMQT oscillations.} 
\end{figure}

\newpage 

\begin{figure} 
\centerline{\psfig{file=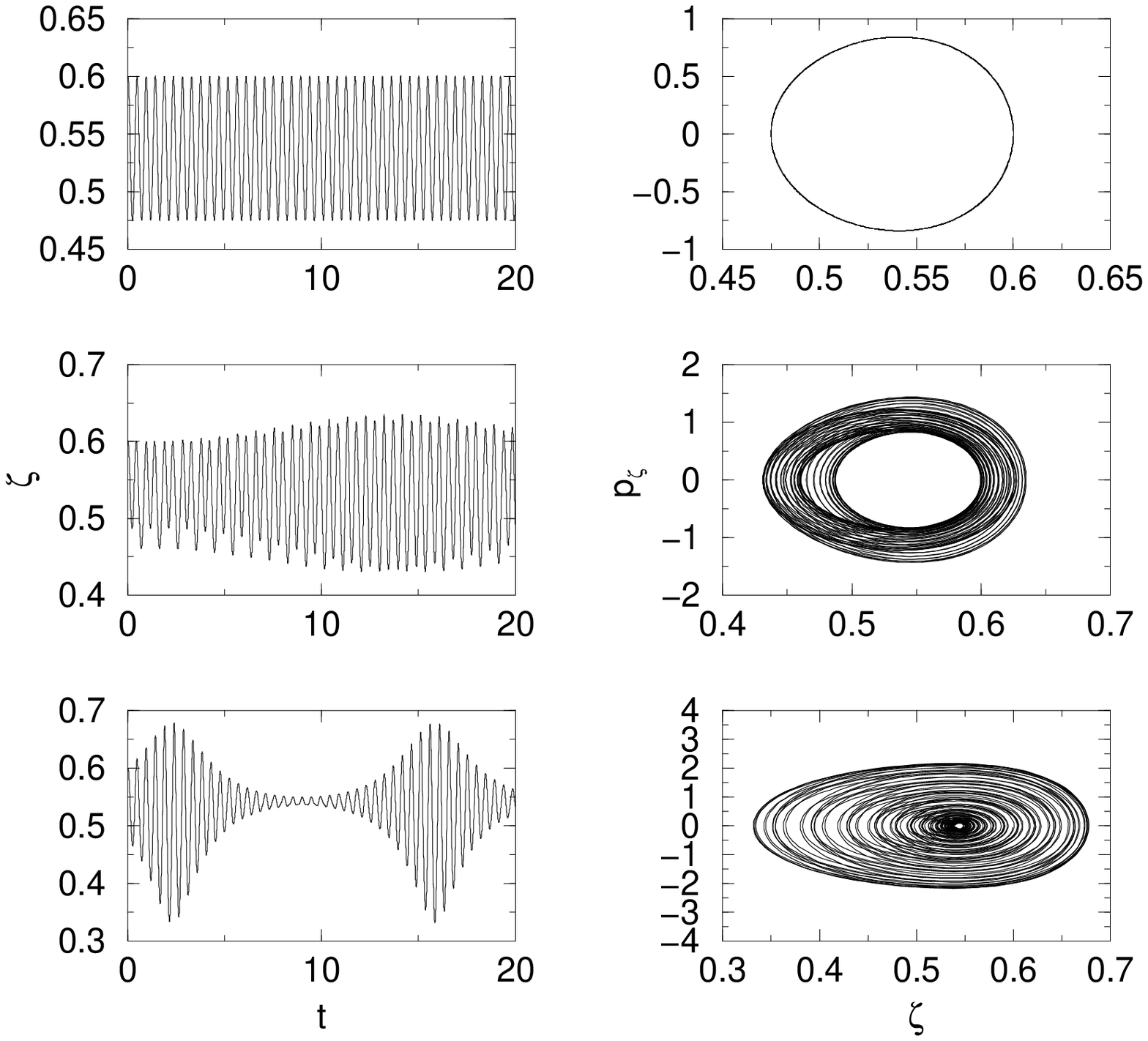,height=5.in}}
\caption{MQST regime. Fractional population 
imbalance $\zeta(t)$ and its phase-space portrait 
$(\zeta,p_{\zeta})$. Initial conditions: 
$\zeta(0)=0.6$, $\phi(0)=0$. 
$\Lambda=\Lambda_0(1+\epsilon \sin{(\omega t)})$ 
where $\Lambda_0=25$. From top to bottom: 
(a) $\epsilon = 0$; (b) $\epsilon = 0.1$ and 
$\omega= \omega_{ST}/2$; (c) $\epsilon =0.1$ 
and $\omega = 2 \omega_{ST}$;  
where $\omega_{ST}$ is the frequency of unperturbed MQST oscillations.} 
\end{figure} 

\newpage

\begin{figure}
\centerline{\psfig{file=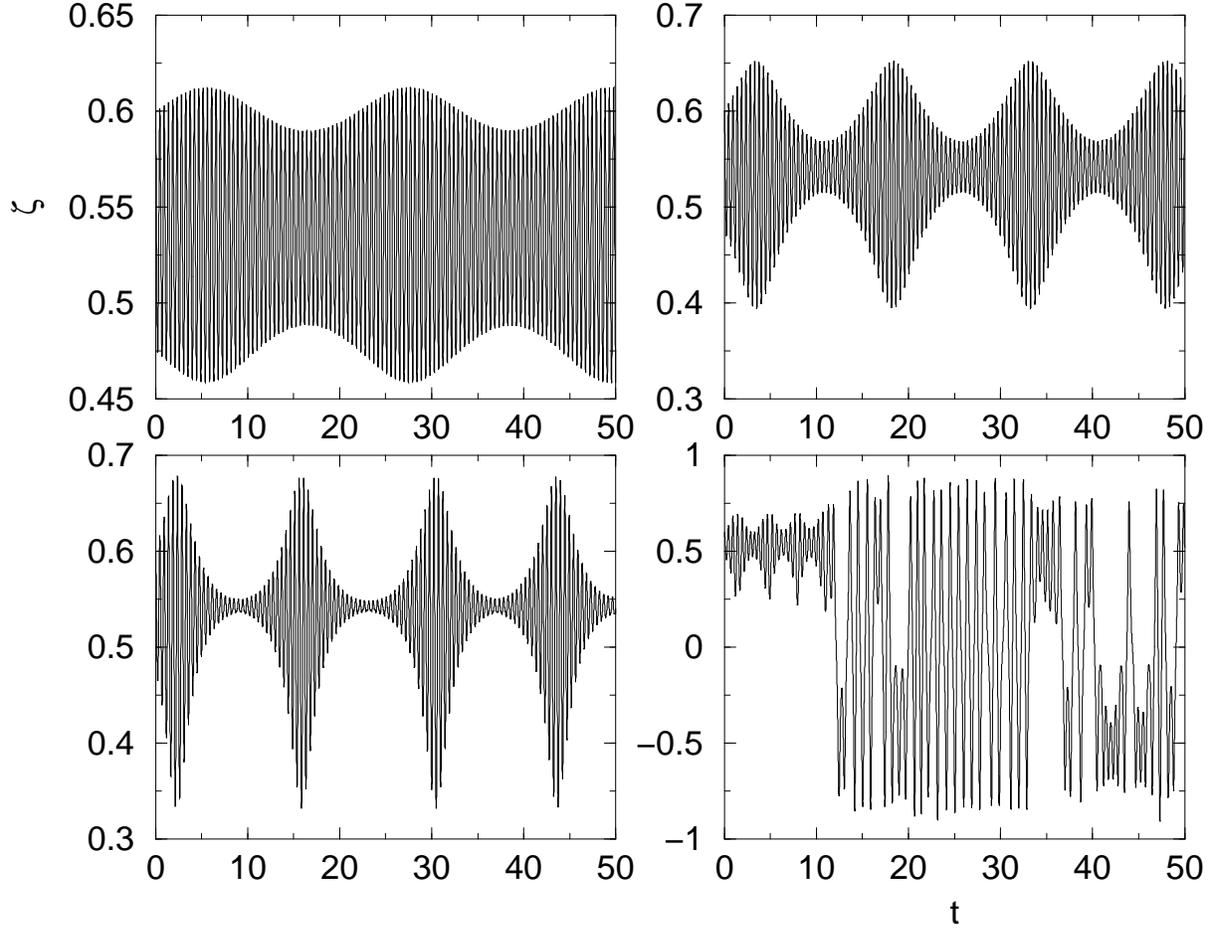,height=5.in}}
\caption{MQST regime at the parametric resonance. 
Fractional population imbalance $\zeta(t)$ of the Bose condensate. 
Initial conditions: $\zeta(0)=0.6$, $\phi(0)=0$. 
$\Lambda=\Lambda_0(1+\epsilon \sin{(\omega t)})$ 
where $\Lambda_0=25$ and $\omega = 2\omega_{ST}$ with 
$\omega_{ST}$ the frequency of unperturbed MQST oscillations.
From top to bottom and from left to right: (a) $\epsilon = 0.01$; 
(b) $\epsilon = 0.05$; (c) $\epsilon = 0.1$; (d) $\epsilon = 0.2$. } 
\end{figure} 
        
\end{document}